\newcommand{\kB}{k_B}
\newcommand{\tB}{\tilde{B}}
\newcommand{\Ffh}{F_{fh}}

\newcommand{\Ffhe}{F_{fhe}}
\newcommand{\chicr}{\chi_{c}}
\newcommand{\Psep}{\mkern-5mu}

\newcommand{\pA}{\phi_A}

\newcommand{\phA}{\phi_{\infty}}
\newcommand{\pB}{\phi_B}

\newcommand{\NB}{N_B}

\newcommand{\NA}{N_A}

\newcommand{\Eqn}[1]{Eq.(\ref{#1})}

\newcommand{\Fig}[1]{Fig.(\ref{#1})}

\documentclass[aps,prl,superscriptaddress,twocolumn,showpacs]{revtex4-1}
\pdfoutput=1
\usepackage{graphicx}
\usepackage{color}
\usepackage[ugly]{nicefrac}
\usepackage{amsmath}
\usepackage{amsthm,color}
\begin{document}


\title{Elasticity dominated surface segregation of small molecules in polymer mixtures}


\author{Jaros{\l}aw Krawczyk}
\affiliation{Department of Mathematical Sciences, Durham University, South Road, Durham DH1 3LE, United Kingdom.}
\affiliation{Department of Molecular Physics, {\L}{\'o}d{\'z} University of Technology, {\.Z}eromskiego 116, 90-924 {\L}{\'o}d{\'z}, Poland}
\author{Salvatore Croce}
\affiliation{Department of Mathematical Sciences, Durham University, South Road, Durham DH1 3LE, United Kingdom.}
\author{T. C. B. McLeish}
\affiliation{Department of Physics, Durham University, South Road, Durham DH1 3LE, United Kingdom.}
\author{Buddhapriya Chakrabarti}
\email{buddhapriya.chakrabarti@durham.ac.uk}
\affiliation{Department of Mathematical Sciences, Durham University, South Road, Durham DH1 3LE, United Kingdom.}

\date{\today}

\begin{abstract}
We study the phenomenon of migration of the small molecular weight component of a binary polymer mixture to the free surface using mean field and self-consistent field theories. By proposing a free energy functional that incorporates polymer-matrix elasticity explicitly, we compute the migrant volume fraction and show that it decreases significantly as the sample rigidity is increased. Estimated values of the bulk modulus suggest that the effect should be observable experimentally for rubber-like materials. This provides a simple way of controlling surface migration in polymer mixtures and can play an important role in industrial formulations, where surface migration often leads to decreased product functionality. 
\end{abstract}


\maketitle

\textit{Introduction:} When a polymer mixture having mobile components of different molecular weights and with an interface that is free to atmosphere is left to equilibrate, the small molecular weight component migrates to the surface~\cite{Jones:book, Budkowski1999, Lipatov2002}. Several industrial formulations, \textit{e.g.} chocolate~\cite{Lonchampt2004}, food packaging~\cite{Bhunia2013} \textit{etc.} suffer from this ubiquitous problem. While many experimental~\cite{Budkowski1999, Geoghegan2003, Jones:book} and theoretical studies~\cite{Binder1995} of this phenomenon exist, a good quantitative agreement between theoretical predictions and experiments is still lacking~\cite{Jones:book}. Further, experimental strategies to control the amount of material migrating to the surface is in a nascent stage of development.

In this letter we ask: {\it How does the elasticity of the polymer matrix influence surface migration of small molecules in polymer mixtures?} We propose a free energy functional that incorporates elasticity of the polymer mixture explicitly, a feature that has been ignored in previous surface segregation studies. Using mean field theory (MFT) and self-consistent field theory (SCFT) we show that as the sample rigidity is increased {\it (i)} the migrant fraction decreases, and {\it (ii)} a wetting transition can be avoided (demonstrated by a geometric construction~\cite{Cahn1977, deGennes1985}). These results are of paramount importance in industrial product formulations where surface migration of small molecular weight component results in decreased functional performance of the product.

\begin{figure}[t]
\includegraphics[width=8cm]{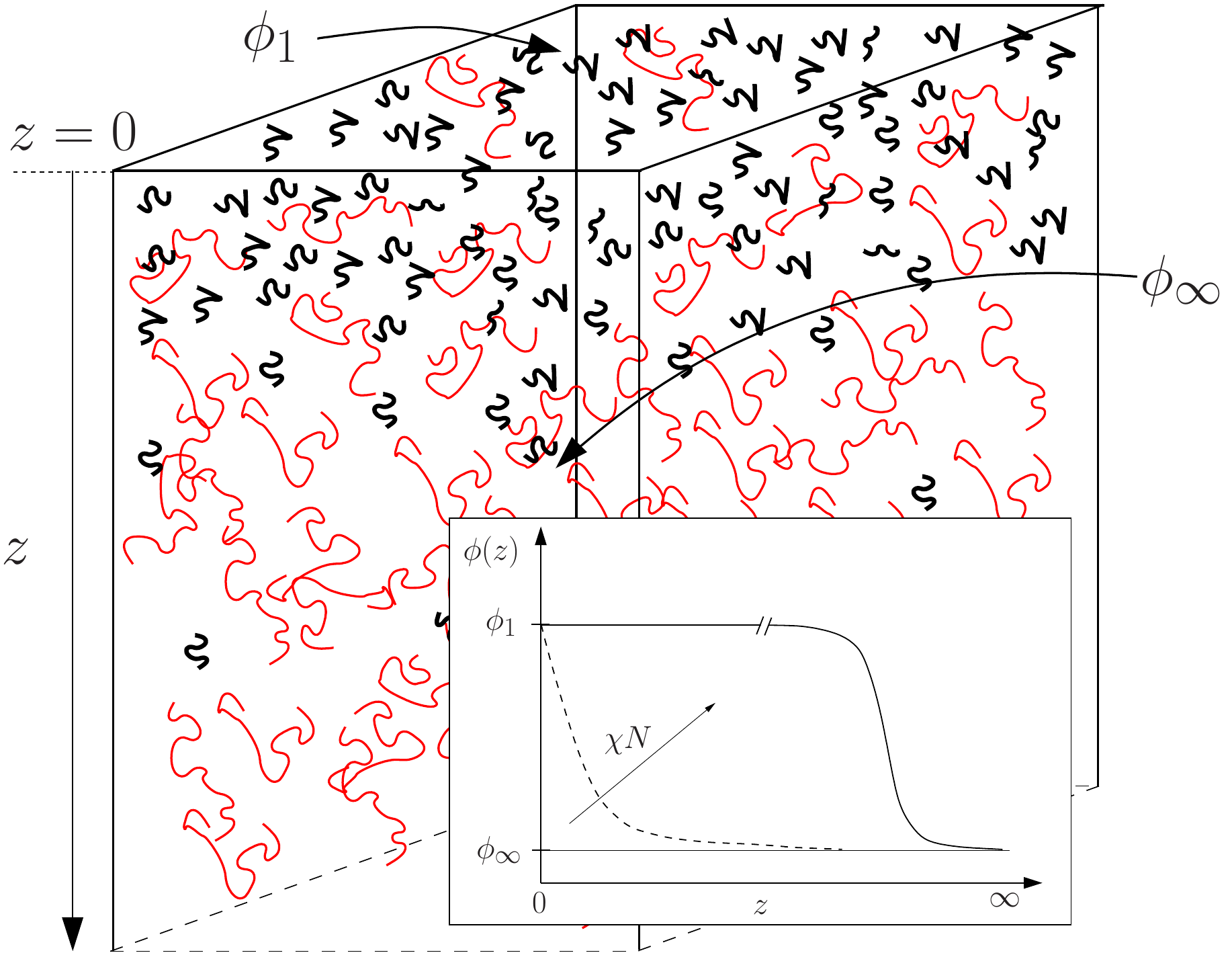} 
\caption{Schematic figure showing a mixture of low (black) and high (red) molecular weight polymers, with the low molecular weight component migrating to the free interface $z=0$. A semi-infinite geometry is assumed. The volume fraction of the migrant in bulk and at the surface is denoted by $\phA$, and $\phi_1$ respectively. Inset shows migrant concentration profiles for different values of $\chi N$. For low values of $\chi N$ a monotonically decreasing concentration profile is observed (dashed line). As $\chi N$ increases a wetting transition, characterised by a macroscopically thick migrant layer (solid line with a break) is observed.
\label{fig01}}
\end{figure}

\textit{Surface Migration:} For a binary mixture, the component with the lower surface energy will migrate to the interface. A balance between loss of translational entropy and gain in surface energy dictates the equilibrium morphology of such systems. This is shown in Fig.~\ref{fig01} with a high migrant (black) concentration close to the interface ($z=0$) of a mixture of low and high (red) molecular weight polymers. The migrant concentration decreases monotonically to the bulk concentration $\phA$ as $z \rightarrow \infty$. A crucial parameter that dictates the thermodynamics of the system is $\chi N$, where $\chi$  is the miscibility parameter, and $N$ the molecular weight of the migrant. As $\chi N$ increases a wetting transition, characterised by a macroscopic wetting layer is observed (Fig.~\ref{fig01} inset)~\cite{Jones:book}.

Surface migration was first observed using X-ray photoemission spectroscopy~\cite{Pan1985} and the resolution of the depth profile of the migrant concentration was improved significantly using neutron-reflectivity~\cite{Jones1990}. Further studies concentrated on the theoretical aspects of the migration by Schmidt and Binder~\cite{Schmidt1985} and subsequently a comparison between theory and experiments~\cite{Jones1989}. The wetting transition of polymer mixtures at the air-mixture interface was first demonstrated by Steiner \textit{et al.}~\cite{Steiner1992}. Experimental and theoretical developments of this phenomenon has recently been reviewed by a few authors~\cite{Budkowski1999, Geoghegan2003}.

\begin{figure}[t]
\includegraphics[width=8cm]{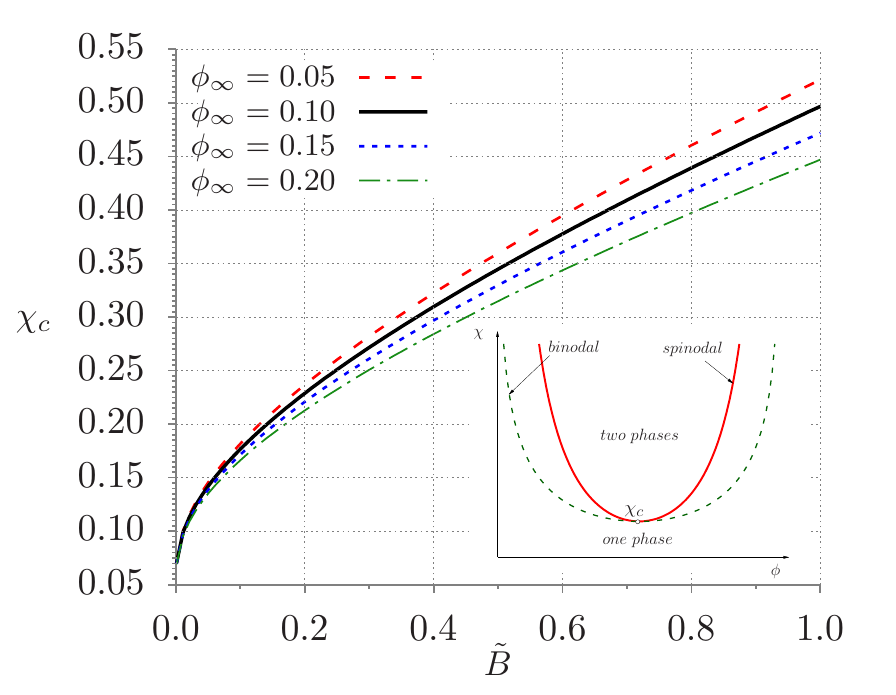} 
\caption{The variation of $\chicr$  on the elastic modulus $\tB$ of a phase separating binary mixture with elastic interactions for different values of migrant volume fraction $\phA$ ($0.05$ (red dashed line), $0.10$ (black solid line), $0.15$ (blue dotted line) and $0.20$ (green dash-dotted line)). Figure shows $\chicr$ increases with $\tB$ (as $\sim \sqrt{\tB}$) indicating that softer systems are more susceptible to phase separation and decreases with $\phA$ for a fixed $\tB$. Inset shows phase diagram of polymer mixtures without elastic interactions. For $\chi < \chicr$ the system remains in one phase. Increasing $\chi$ beyond $\chicr$ phase separation occurs. The coexistence (green dashed line) and spinodal (red solid line) curves demarcating phase boundaries is shown. 
\label{fig02}}
\end{figure}

\textit{Flory-Huggins theory:} The thermodynamics of mixing of two chemically different polymers is well described by Flory-Huggins (FH) theory~\cite{Rubinstein:book}. The mixing free energy per unit volume
is given by
\begin{equation} 
\frac{\Ffh[\phi]}{\kB T}=\frac{(1-\phi)}{\NB} \log(1-\phi)+\frac{\phi}{\NA} \log(\phi)+\chi\phi(1-\phi),
  \label{Ffh}
\end{equation}
where $\chi$ is the miscibility parameter, and $\NA$, and $\NB$ are the degree of polymerisation of A and B polymers respectively. The volume fractions of the A ($\pA = \phi$), and B ($\pB = 1 - \phi$) polymers in Eq.~\ref{Ffh} thus satisfy the incompressibility constraint $\pA + \pB=1$. The phase behaviour of such systems is well known~\cite{Rubinstein:book}. Below a critical value of the miscibility parameter $\chi < \chi_{c}=1/(2\NA)+1/(2\NB)+1/(\sqrt{\NA\NB})$ the equilibrium phase is a homogeneous mixture of A and B polymers. For $\chi > \chi_{c}$ (\textit{e.g.} effected by changing temperature) phase segregation occurs with the system separating into A and B rich regions. Depending on the parameters, a first or second order transition might be observed. This is schematically shown in the inset of Fig.~\ref{fig02} (solid line).

\textit{Schmidt-Binder formalism:} While FH free energy describes the phase separation process in bulk it cannot be directly applied to study segregation close to an interface that is exposed to atmosphere. Cahn's~\cite{Cahn1977} seminal work provides a cue in this case. This framework offers a way of calculating the concentration profile of a fluid near a wall, given a limiting fluid concentration, using the calculus of variations. The Flory-Huggins as well as Cahn's theory have successfully been combined into a single mean field description to describe the surface segregation of binary polymer mixtures by Schmidt and Binder~\cite{Schmidt1985} (referred as SB henceforth). The SB free energy functional for a semi-infinite system ($z>0$) with an attractive surface having area $A$ at $z=0$ is given by
\begin{equation}
\frac{F_{SB}[\phi]}{A\kB T}=
\int_{0}^{\infty}dz \left\{\frac{\Ffh[\phi]}{\kB T}+k(\phi)\left(\frac{d \phi}{dz}\right)^2-\Delta \mu \phi\right\}+ F_s(\phi_1),
\label{FSB}
\end{equation}
where $k(\phi) = \frac{a^{2}}{36 \phi (1-\phi)}$ is the coefficient associated with the energetic cost of creating a concentration gradient (obtained within the random phase approximation~\cite{deGennes:book,Jones1989,Schmidt1985}), and $\Delta \mu$ models the exchange chemical potential. The SB functional also incorporates the surface free energy gain of the migrant $F_s(\phi_1)$ expressed as a polynomial expansion of the migrant volume fraction at the surface, ($\phi_1 = \phi(z=0)$) and is given by $F_s(\phi_1)=-\phi_1\mu_1-\frac{g}2\phi_1^2$, where $\mu_1$ is the surface chemical potential and the coefficient $g$ characterises the change in bulk interactions due to the surface~\cite{Schmidt1985,Jones1993}. Within the gamut of square gradient theory the free energy functional in \Eqn{FSB} can be minimised $\delta F_{SB}[\phi]/\delta \phi = 0$, to yield an integral expression for $z(\phi)$, which can be inverted to obtain the concentration profile of the migrant $\phi(z)$~\cite{Schmidt1985}. For small values of $\chi N$ an exponentially decaying profile shows reasonable agreement with experimental data~\cite{Jones:book}.

\textit{Elastic Flory-Huggins theory:} We now explore the role of polymer matrix elasticity in the small molecule migration process. If one component (B polymer in our case) forms an elastic network as in cross-linked gels (reticulated permanent network) or transient network (as in polymer solutions) then its entropic contribution to the FH mixing free energy would be negligible in comparison to that of the migrant. Assuming Flory-Rehner form of free energy~\cite{FloryRehner1943} describing the energy cost of a migrating oligomer as it pushes its way through the matrix, the Flory Huggins elastic free energy $\Ffhe$ can be written as
\begin{equation} 
\frac{\Ffhe}{\kB T}=\frac{\phi\log(\phi)}{\NA}+\chi\phi(1-\phi)+\frac{F_{el}}{\kB T},
\label{Ffhel} 
\end{equation}
where $F_{el}=\tB \frac{n}{2}(\lambda^2+\frac{2}{\lambda}-3)$, modelling uniaxial network deformation~\cite{Rubinstein:book}, with $\lambda$ representing the relative chain extension,  $n$ the number of chains in the network and $\tB$ the elastic modulus. The relative chain extension is defined as $\lambda=R/R_0$, where $R_0$ and $R$ denote the length of the polymer before and after deformation respectively. Assuming the volume fraction of the matrix polymer before and after deformation being $\phi_{B_0}=V_{B_0}/V$, and $\pB=V_B/V$ respectively, we obtain an expression of the relative chain extension $\lambda$ in terms of the volume fractions, \textit{i.e.} $\pB/\phi_{B_0}=V_B/V_{B_0}=(R/R_0)^3=\lambda^{3}$. Since $\phi_{B_0}=1-\phi_{A_0}=1-\phA$ and $\pB=1-\phi$, $\lambda=[(1-\phi)/(1-\phA)]^{\nicefrac{1}{3}}$. The volume fraction of migrant molecules in the sample is $\phi$. Following the theory of elastic rubber networks~\cite{Treloar:book}, the number of chains in the network can be estimated as $n = \NB/V = (1 - \phA)$.  The number of chains that form the polymer matrix 
The elastic free energy in \Eqn{Ffhel} is thus
\begin{equation}
F_{el}=\tB\frac{(1-\phA)}{2}\Bigg[\left(\frac{1-\phi}{1-\phA}\right)^{\Psep\nicefrac{2}{3}} +2 \left(\frac{1-\phA}{1-\phi}\right)^{\Psep\nicefrac{1}{3}} -3\Bigg],
\label{Eq_Fel2}
\end{equation}
where $\tilde{B}$ is the elastic modulus expressed in units of $k_{B} T$. The free energy that describes the small molecule migration through a matrix where elastic effects have been explicitly incorporated is therefore given by
\begin{equation}
\frac{F_{tot}[\phi]}{A\kB T}=
\int_{0}^{\infty}dz\left\{ \frac{F_{fhe}}{\kB T}+ k(\phi)\left(\frac{d \phi}{dz}\right)^2 -\Delta \mu \phi\right\}+F_s(\phi_1),
\label{Ftot}
\end{equation}
where $F_{fhe}$ is the elastic Flory Huggins functional in \Eqn{Ffhel} and the gradient, exchange chemical potential and surface contributions to the free energy is the same as the SB free energy functional in \Eqn{FSB}.

The role of elasticity in the phase separation of binary polymer mixtures where both species are cross-linked have been investigated earlier~\cite{deGennes1979,Read1995,Clarke1997}. Such a system shows microphase separation and is different from the functional proposed here (\Eqn{Ftot}). 

Before discussing the surface segregation process we consider the bulk thermodynamic behaviour of the system described by \Eqn{Ffhel}. This can be obtained easily by minimising the elastic FH free energy with respect to $\phi$. The minimisation procedure leads to a relation between $\phi$ and $\chi$, which for bulk concentration $\phA$ corresponds to the binodal curve $\chi=[1-\log(\phA)-\NA\Delta\mu]/[\NA(1-2\phA)]$. It is interesting to note that $\chi$ parameter does not depend on the elastic modulus $\tB$. The critical value of $\chi_{c}$ above which the mixed phase is unstable, obtained from the relation $\partial^3 \Ffhe/\partial\phi^3 = 0$, however increases with increasing $\tB$. This is shown in \Fig{fig02} with $\chi_c \sim \sqrt{\tB}$ for different values of $\phA$. As shown in \Fig{fig02} $\chicr$ decreases with increasing $\phA$ for a fixed $\tB$. This can be understood as follows. As the volume fraction of the migrant increases, the available free volume decreases and hence entropy decreases. Since a balance between entropic and enthalpic contributions dictates the equilibrium, a lower value of enthalpy (and hence lower $\chi$) is required to bring about the phase separation. With $\chicr$ increasing with $\tB$ the single phase region of a rigid system is stable for larger values of $\chi$ in comparison to polymer mixtures without elastic interactions. The phase behaviour of the binary polymer mixture without matrix elasticity is shown in the inset of \Fig{fig02}.

\begin{figure}[ht]
\includegraphics[width=8cm]{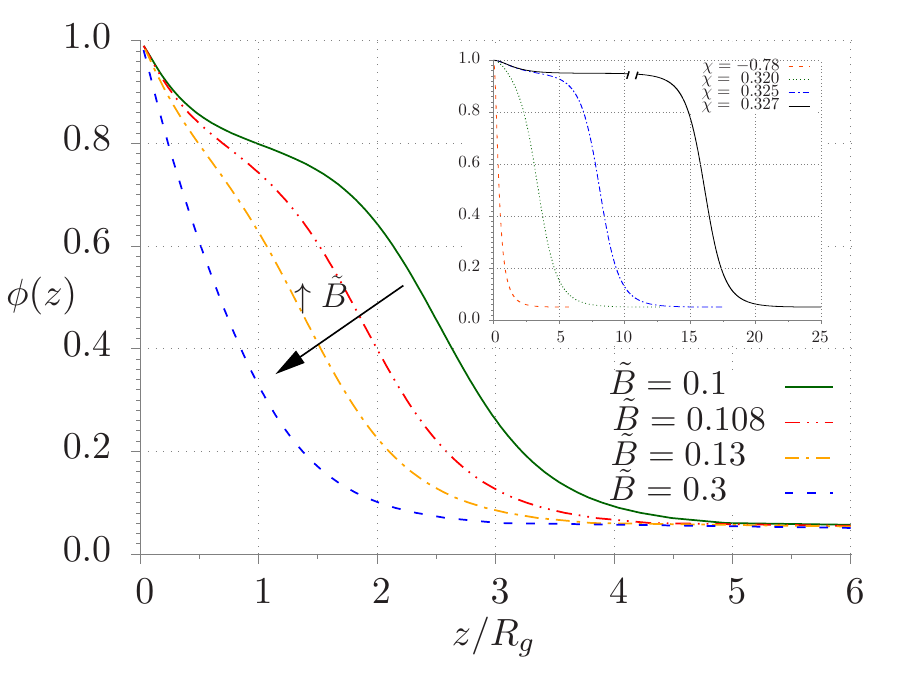}
\caption{Migrant concentration profiles $\phi(z)$ for the SB model including the elasticity obtained by minimising \Eqn{Ftot} for ($\chi=0.320$ and $\NA=10$) and increasing $\tB$ ($0.1$ (green solid line), $0.108$ (red dash double dotted line), $0.13$ (yellow dash dotted line), $0.3$ (blue dashed line)). A wetting transition is not observed in this model. Inset, shows concentration profiles for SB model without elasticity for symmetric case $\NA=\NB=10$ for increasing $\chi$ ($-0.78$ (red dashed line), $0.320$ (green dotted line), $0.325$ (blue dash dotted line) and $0.327$ (black solid line with a break) indicating the formation of a macroscopic wetting layer. 
\label{fig03}}
\end{figure}

\textit{Surface segregation for elastic FH theory:} The SB formalism, outlined earlier can be used to compute the concentration profile of the migrant $\phi(z)$ close to the interface for the phenomenological free energy functional described by \Eqn{Ftot}. \Fig{fig03} shows migrant concentration profiles for both systems, a symmetric binary polymer mixture having a bulk concentration $\phA = 0.05$ with and without elastic interactions. The inset shows $\phi(z)$ as a function of depth $z$ for different values of $\chi$ for $\NA = \NB = 10$ in the absence of elasticity (obtained by minimising \Eqn{FSB}). For smaller values of $\chi$ ($-0.78$ (red dashed line)) an approximate exponentially decaying profile is observed. As $\chi$ increases, migrant concentration reaching the surface increases monotonically ($\chi = 0.320, 0.325$) and beyond a critical value $\chicr = 0.327$ a macroscopic wetting layer is observed. In contrast, when elastic interactions are included (main panel \Fig{fig03}), the migrant fraction for the same value of miscibility parameter $\chi$ ($0.320$), obtained by integrating the area under the curve $\phi(z)$ decreases monotonically with increasing $\tB$. For lower values of the modulus, $\tB = 0.1, 0.108$ a shoulder (reminiscent of a rounded wetting transition) is observed. For higher values of $\tB$ ($0.13$, $0.3$) an exponentially decaying profile is obtained suggesting elastic interactions severely inhibiting migration.

While physically intuitive and relatively straightforward to implement the SB model has some disadvantages. First, the surface migrant fraction $\phi_1$ is an additional input and cannot be calculated from the model. In order to establish our main result, namely, elastic interactions inhibit surface migration as the matrix rigidity is increased, we employ a self-consistent field theoretic approach where this limitation does not exist. However, both the SB model and the SCFT framework suffers from the limitation that the bulk volume fraction $\phA$ is held constant, no matter how much material flows to the surface. Modifications to the SB and SCFT framework that do not suffer from this drawback will be reported elsewhere~\cite{self}.

\textit{Self-Consistent Field Theory:} First introduced in context of polymers by Edwards~\cite{Edwards1965}, self-consistent field theory (SCFT) has been successfully employed to solve equilibrium behaviour of  polymeric systems~\cite{Fredrickson:book}. We employ the SCFT formalism developed for end absorbed polymer brushes in polymer matrices~\cite{Shull1990,Shull1991} and adapt it to our situation. 

We consider two types of chains A, and B, with each component interacting via a mean-field potential on a lattice. The configuration of a polymer chain on the lattice can be visualised as a random walk with the variable representing a segment, akin to ``time''. The distribution function of a chain of a given species $q_{k}(z, t)$, (where $k=A,B$ indicates the species) in the absence of interaction is Gaussian, satisfying a diffusion equation with the diffusion constant given by the squared radius of gyration $R_{g} = N a^{2}/6$. When the chains interact, $q_{k}(z, t)$ satisfies the modified diffusion equation 
\begin{equation}
\label{diffu}
\frac{\partial q_{k}(z,t)}{\partial t} = N\frac{a^2}{6} \nabla^2 q_{k}(z,t) - w_{k}(z) q_{k}(z, t), 
\end{equation}
where $a$ is the Kuhn length, $N$ the degree of polymerisation and $w_{k}(z)$ is a mean field which takes into account the interactions between the monomer and its neighbouring chains of either species. Since we consider segregation in one dimension the variable $z$ in \Eqn{diffu} represents the distance from the surface, while ``time'' $t$ indicates a particular segment of a chain. The mean field potential for the migrant species $w_{A}(z)$ is related to it's chemical potential by
\begin{equation}
\label{mua}
w_{A}(z) = \frac{1}{\NA} \left( \mu^{0}_{A}(z) - k_{B} T \log \phi_{A}(z) \right),
\end{equation} 
with the bare chemical potential $\mu^{0}_{A}(z) = \NA \Ffhe + \NA \phi_B \left(  \partial \Ffhe/ \partial \phi_A - \partial \Ffhe/ \partial \phi_B \right)$. For the matrix, we neglect the entropic contribution in \Eqn{mua}. Thus the mean field potential $w_{B}(z)$ is the same as the bare chemical potential $\mu^{0}_{b}$, which has the same functional form as $\mu^{0}_{A}$ with A and B indices interchanged. The volume fractions $\phi_{k}$ can be computed from the distribution function using
 \begin{equation}
\label{phisc}
\phi_{k}(z)=\frac{e^{\beta \mu_{k}(z)}}{N_{k}} \int_{0}^{N_{k}} dt  \; \; q_{k}(z,t)q_{k}(z,N_{k}-t),
\end{equation}
where $\mu_{k}(z) = \mu^{0}_{k}(z) + F_{s}$ for the matrix and the migrant, with $F_{s}$ (defined in units of $1/k_B T$) being the surface free energy. \Eqn{diffu}-\Eqn{phisc} are solved self-consistently~\cite{Shull1990,Shull1991} with the boundary conditions at the surface $q_{A}(z=0,t) = e^{-F_s}$, and $q_{B}(z=0,t) = 0$ for the migrant and the matrix respectively. The initial conditions $q_{k}(z,t=0) = 1$ is used for both species.

\begin{figure}[ht]
\includegraphics[width=8cm]{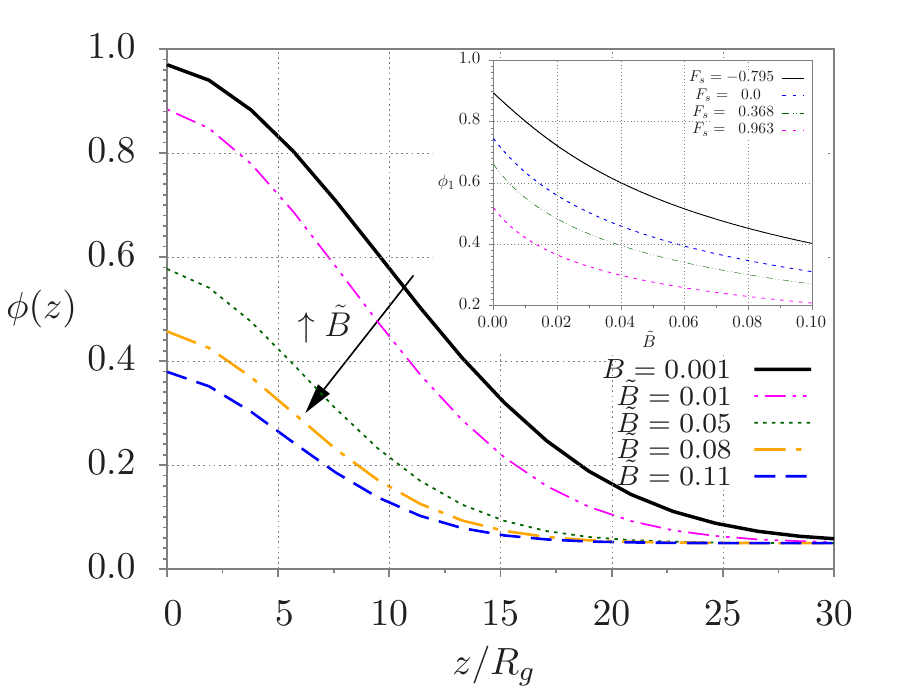}
\caption{Migrant concentration profiles $\phi(z)$ for different elastic modulii $\tB$ of the polymer matrix ($0.001$, (black solid line), $0.01$ (magenta dash double dotted line), $0.05$ (green dotted line), $0.08$ (yellow dash dotted line), and $0.11$ (blue dashed line)). The amount of material flowing to the surface decreases with increasing $\tB$.  The dependence of the surface fraction $\phi_1$ as a function of $\tB$ for different surface free energy $F_s$ ($-0.795$ (black solid line), $0.0$ (blue dotted line), $0.368$ (green dash dotted line), and $0.963$ (magenta dashed line)) is shown in the inset. As expected the volume fraction decreases for system with higher $F_s$.
\label{fig04}}
\end{figure}

The concentration profile of the migrant as a function of distance from the surface (in units of $R_g$) obtained from the SCFT calculation is shown in \Fig{fig04} for a miscibility parameter $\chi=0.22$ and surface energy $F_{s} = -2.0$. The migrant polymer has a Kuhn length $a=1$ and a degree of polymerisation $\NA=10$. As the elastic modulus of the matrix $\tB$ is increased (from $0.001$ to $0.11$) the amount of material migrating to the surface decreases monotonically. In contrast to the SB model where $\phi_1$ is an additional input, (($\phi_1 = 1.0$) in \Fig{fig03}), it can be calculated within the SCFT framework. \Fig{fig04} shows $\phi_1$ decreasing monotonically with increasing $\tB$. The inset shows \Fig{fig04} the variation of the migrant concentration at the surface $\phi_1$ as function of $\tB$ for different values of the surface energy $F_s$. The effect of elasticity on the migrant fraction $\phi_1$ is more pronounced for low values of $\tB$, ($\approx 0 - 0.02$). As expected $\phi_1$ decreases with increasing surface free energy $F_s$ for a given value of $\tB$. For the elastic systems considered here, a wetting transition is not observed. A direct comparison between the parameters in SB model and a variant of the SCFT method presented here~\cite{Genzer1994} is currently underway.

\begin{figure}[t]
\includegraphics[width=8cm]{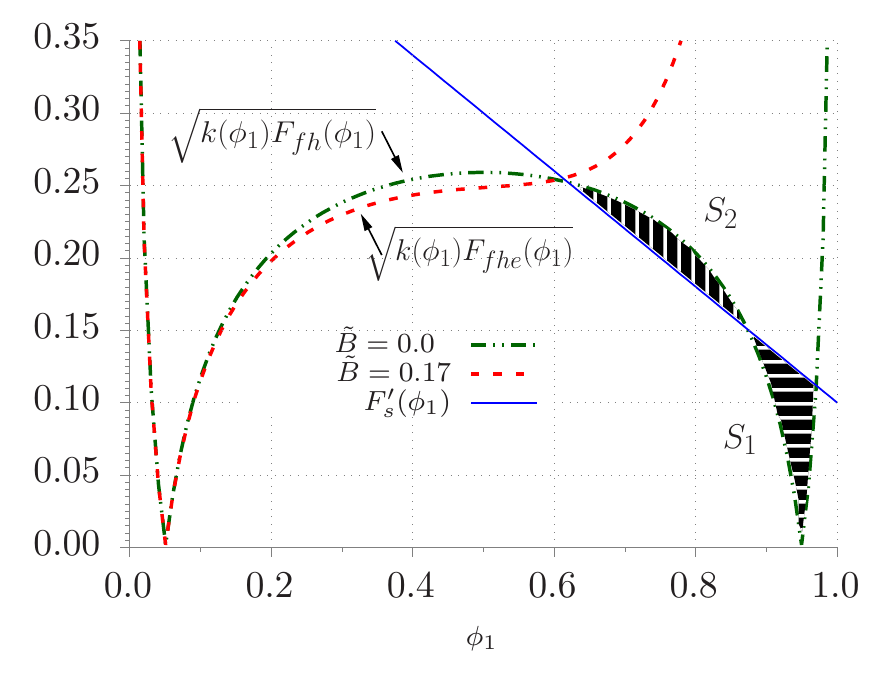} 
\caption{Cahn construction showing first order wetting transition for the Flory Huggins free energy functional, $\Ffh$ (\Eqn{Ffh}) (green dash dotted line).  An intersection between $\Ffh(\phi_1)$ and $F^{\prime}_{s}(\phi_1)$ (blue solid line) at three points intersects demarcating areas $S_1$ and $S_2$, such that $S_1 > S_2$ indicates a first order wetting transition. A similar graphical construction for the elastic Flory-Huggins functional $\Ffhe$ (\Eqn{Ffhel})) with $\tB = 0.17$ (red dashed line) shows one intersection, indicating the absence of wetting transition.\label{fig05}}
\end{figure}

\textit{Cahn construction:} A geometric way of demonstrating the absence of a wetting transition has been proposed by Cahn~\cite{Cahn1977,deGennes1985} and applied in context of binary polymer mixtures~\cite{Budkowski1997}. A calculation of the surface migrant concentration $\phi_1$ involves solving the equation
\begin{equation}
\label{Cahn}
F^{\prime}_{s}(\phi_1) = \sqrt{k(\phi_1) F(\phi_1)}, 
\end{equation}
where $k(\phi)$ has the same meaning as \Eqn{Ftot}, and $F(\phi_1)$ refers to the $\Ffh$ for SB model and $\Ffhe$ when elastic interactions are present. A graphical method of solving \Eqn{Cahn} is shown in \Fig{fig05}, plotting the surface $F^{\prime}_{s}(\phi_1)$ (blue solid line) for $\mu_1 = -0.5$, and $g = 0.4$, and bulk free energy contributions $\sqrt{k(\phi_1) F(\phi_1)}$ as a function of $\phi_1$ for a system with (red dashed line) and without (green dash dotted line) elastic interactions. In the absence of elasticity $\tB = 0$ the curves intersect at three points, demarcating areas $S_1$ and $S_2$ such that $S_1 > S_2$ indicating a wetting transition. For a finite value of $\tB$ ($0.17$ in \Fig{fig05}) the wetting transition is absent~\cite{Cahn1977, deGennes1985}. 

\textit{Conclusion:} In conclusion, we have analysed the role of matrix elasticity on the surface migration of small molecules in binary polymer mixtures using mean field and self-consistent field theories. We have shown that increasing the rigidity of the matrix leads to significant reduction of the migrant fraction on the surface. Further, by increasing the elastic modulus of the polymer matrix a wetting transition can be avoided for systems having miscibility parameters in the range that would otherwise have led to it. This provides a novel way of controlling surface migration in complex industrial formulations  such as adhesives in hygiene products where surface migration leads to decreased product functionality. To the best of our knowledge, the only experimental system (despite significant differences) related to the theory presented here investigates segregation processes in polystyrene networks~\cite{Geoghegan2000}. We hope that our theoretical work will prompt experimental studies in this direction.

\begin{acknowledgments}
\textit{Acknowledgements:} JK, TCBM, and BC thank Procter and Gamble for financial support. SC and BC acknowledge funding via the ERC Marie Curie ITN MICSED grant under FP7 program. The authors thank Richard Thompson, Elise Sabattie, Mark Wilson and Jos Tasche from Durham University, Marc Hamm from Henkel, and Todd Mansfield, Gabriela Schaefer, Torsten Lindner and Mattias Schmidt from P\&G for helpful discussions. 
\end{acknowledgments}


\end{document}